\documentstyle[prd,aps]{revtex}
\begin{document}
\def\be{\begin{equation}}
\def\ee{\end{equation}}

\title
{On the detection of scalar hair}

\author{
Narayan Banerjee\footnote{Email address:narayan@juphys.ernet.in}} 
\address{Relativity and Cosmology
Research Centre, Department of Physics, Jadavpur University, 
Calcutta-700032, India}
\author{
Somasri Sen\footnote{Email address: somasri@mri.ernet.in}}
\address{Mehta Research Institute}
\author{
Naresh Dadhich\footnote{Email address: nkd@iucaa.ernet.in}}
\address{Inter University Centre for Astronomy and Astrophysics, Post
Bag 4, Ganeshkhind, Pune 411007, India}
\maketitle

\begin{abstract}
It is shown that the conclusion regarding the existence of a scalar
hair for a black hole in a nonminimally coupled self interacting scalar 
tensor theory can be drawn from the that for a scalar field minimally
coupled to gravity by means of a conformal transformation.
\end{abstract}

\section{introduction}

The general belief that a black hole cannot have a scalar hair was
disproved when it had been shown that in dilaton gravity, where a scalar
field is nonminimally coupled to curvature, a black hole may indeed have a
scalar hair \cite{garfinkle}.  This result motivated a lot of workers towards 
the search for the existence of a scalar hair in other nonminimally coupled
scalar tensor theories of gravity. As there is no rigorous theorem  to suggest in
which cases a scalar hair might exist, the usual practice is to look at different
examples.  Unfortunately it is often quite difficult to find an exact solution for
the spacetime metric with a nonminimally coupled scalar field.  The exact solutions
for a minimally coupled scalar field, on the other hand, are more readily found as
the field equations are far less involved. Saa \cite{saa} proved a very useful
theorem which states that the conclusions regarding the existence of a scalar hair
for a wide class of nonminimally coupled scalar tensor theories can actually be
drawn from the solutions for the metric for a spacetime with a scalar field
minimally coupled to gravity. Saa's method proves to be extremely useful as it
covers a very wide class of scalar tensor theories. This method had been generalized
to include an elctromagnetic field \cite{banerjee} and also to find a counterexample
of the no hair conjecture in an axially symmetric black hole where the scalar field
is anisotropic and the spacetime is not asymptotically flat\cite{sen}. Although
Saa's theorem is very general, it does not incorporate a scalar field with a
self-interaction. Mayo and Bekenstein \cite{mayo} included self interacting scalar
fields, i.e. a scalar field with a potential, where the coupling between the
curvature $(R)$ and the scalar field $(\phi)$ has the form $\eta \phi^{2}
R$. \\ 

In this work we generalize Saa' work to include an electromagntic field or any
$U(1)$ gauge field which has a trace free  $(T=T^{\mu}_{\mu}= 0)$  energy
momentum tensor and a self interaction term for the  non minimally coupled 
scalar field. It is also generalization of the Mayo-Bekenstein case for we keep
the coupling between scalar field and curvature free. We have been able
to find a suitable redefinition of scalar field for the generalized
framework as well and then the minimally and non minimally coupled
scaler fields are described by the metrics that are conformally
related. \\
In the next section we shall consider this
transformation and demonstrate how it works which would be followed by a
specific example in section III. We conclude with a discussion in section
IV. 

\section{conformal transformation}

We take a very general action of the form

\be
S = \int{\sqrt{-g}d^{4}x[f(\phi)R - h(\phi)\phi^{,\alpha}\phi_{,\alpha} +
V(\phi)
-F^{\alpha \beta}F_{\alpha \beta}]},
\label{action1}
\ee
where $f(\phi)$, $h(\phi),$ and $V(\phi)$  are functions of the scalar
field  $(\phi)$. $F_{\mu\nu}$ is the antisymmetric  Maxwell tensor. The action is
more general than that used by
Saa\cite{saa} in the sense that it includes a potential $V(\phi)$ and an
elctromagnetic field. The scalar field couples with curvature as $f(\phi)R$, i.e.,
in a similar way as that used in Saa's ansatz and is more general than that used by
Mayo and Bekenstein, where $\phi$ and $R$ couple as $\eta\phi^{2} R$, i.e., only 
with a quadratic $\phi$.\\
The Einstein field equations with the action (\ref{action1}) are,

\be
f(\phi)R_{\mu\nu} - h(\phi)\phi_{,\mu}\phi_{,\nu} -f_{,\mu
;\nu}-\frac{1}{2}g_{\mu\nu}\Box{f} - \frac{1}{2}Vg_{\mu\nu} =T_{\mu\nu},
\label{fe1}
\ee
where $T_{\mu\nu}$'s are the components of the energy momentum tensor
for the electromagnetic field. We use the unit where $c=1$ and $8\pi G=1$, $G$
being the Newtonian constant of gravitation. The wave equation for the scalar field is
\be
2h\Box{\phi} + h' (\phi
)\phi^{,\alpha}\phi_{,\alpha} +f' (\phi) R +V' (\phi)=0.
\label{we1}
\ee
where a prime indicates differentiation with respect to $\phi$.
This kind of non minimally coupled scalar field theories with a self
interaction are very widely used at present in connection with the
quintessence problem, i.e., to build up a cosmological model where the universe
undergoes a late time acceleration\cite{bertolami}. From equation (\ref{fe1}), the
expression for the Ricci scalar $R$ can be written as 
\be
R=\frac{h}{f} \phi^{,\alpha}\phi_{,\alpha} +3\frac{\Box{f}}{f}-\frac{2V}{f}.
\label{R1}
\ee

Now if we effect a conformal transformation of the form
\be
\bar{g}_{\mu\nu}= \Omega^{2} g_{\mu\nu},
\label{trans1}
\ee
where 
\[
\Omega^{2}= f(\phi),
\]
and redefine the scalar field as
\be
\psi(\phi)=\sqrt{2}\int_{\phi_{0}}^{\phi}{d\zeta}\sqrt{\frac{3}{2}(\frac{d}{d\zeta}ln 
f(\zeta))^{2} +\frac{h(\zeta)}{f(\zeta)}},
\label{trans2}
\ee 
the action (\ref{action1}) takes the form
\be
\bar{S}= \int d^{4}x\sqrt{-\bar g}[\bar{R}-\frac{1}{2}\psi^{,\alpha}\psi_{,\alpha}+\bar{V}(\psi) 
-\bar{F}^{\mu\nu}\bar{F}_{\mu\nu}].
\label{action2}
\ee
Here we have written
\be
\bar{V}=\frac{V}{f^{2}}.
\label{pot1}
\ee
An overhead bar indicates quantities in the transformed version. The action 
(\ref{action2}) clearly represents the Einstein-Maxwell along with a minimally
coupled scalar field $\psi$ including a potential $\bar{V}$. The field equations become
\be
\bar{R}_{\mu\nu} =\frac{1}{2}\psi^{,\alpha}\psi_{,\alpha}-
\frac{1}{2}\bar{g}_{\mu\nu}\bar{V} +\bar{T}_{\mu\nu},
\label{fe2}
\ee
where $\bar{T}_{\mu\nu}$ represents the energy momentum tensor due to the
transformed electromagnetic field. It should be noted that
$\sqrt{-g}F^{2}=\sqrt{-g}F^{\mu\nu}F_{\mu\nu}$ is invariant under this conformal
transformation of the metric. The wave equation for the scalar field in this version
looks like
\be
\bar{\Box}\psi + \frac{d \bar{V}}{d\psi} = 0,
\label{we2}
\ee
 The expression of Ricci scalar, followed from equation (\ref{fe2}), in the transformed version is
\be
\bar{R} = \frac{1}{2}\psi^{,\alpha}\psi_{,\alpha} - 2\bar{V}.
\label{R2}
\ee
As the energy momentum tensor of the Maxwell field is trace free, its contribution
does not appear in the expressions for the Ricci scalars (\ref{R1}) and 
({\ref{R2}) explicitly. The electromagnetic field enters into these scalars only
through the solutions for the metric tensor.   \\

To determine the existence of a scalar hair ( or any hair for that matter) for a
black hole, one has to identify the site of the horizon and then see whether this
horizon is regular with a non trivial scalar field ( or the field corresponding to
the particular kind of hair one is looking for). It is obvious that the field
equations for a nonminimally coupled scalar field (\ref{fe1}) are extremely
involved, whereas the eqautions for a minimally coupled one (\ref{fe2}) are much
more amenable. Now if the solution for the metric $\bar{g}_{\mu\nu}$ with a
minimally coupled scalar field with a potential $\bar{V}$ is known, the solution for
the metric in any nonminimally coupled scalar tensor theory can be generated by the
conformal transformation (\ref{trans1}).  \\

As these two metrics are conformally related, the site of the horizon, if there is
any, will remain the same. For a given set of solutions for $\bar{g}_{\mu\nu}$ and
$\psi$, the corresponding solutions and hence the nature of the surface designated
to be the horizon for  a nonminimally coupled theory with known $h(\phi)$ and
$f(\phi)$ can be determined. One must however be careful that the potentials in
these two cases are different, as given by the equation (\ref{pot1}). 
It is naturally expected that with some regularity conditions on $V, \phi, f,h,$ and 
their powers, at the horizon, the singularity structure for the curvature scalar $R$
and $\bar{R}$ would be the same. In other words, the presence or absence of a scalar 
hair for a nonminimally coupled scalar field with a potential $V$ should be the same
as that of a minimally coupled scalar field with a potential $\bar{V}$ given by the
equation (\ref{pot1}). Let us look at this with a Brans-Dicke type of 
coupling.\\ 

If we choose 
\be
f(\phi)=\phi ; \hspace{10mm}    h(\phi)=\frac{\omega}{\phi}, 
\label{bd1}
\ee
the action (\ref{action1}) represents the Brans-Dicke action\cite{brans} with
an electromagnetic field and a self interaction term for the Brans--Dicke scalar field in
the form $V(\phi)$. Without the Maxwell field, this is exactly the action used 
by many authors in order to obtain a reasonable accelerating model of the
universe  (see ref.\cite{bertolami}). With this choice, the expression for
the Ricci scalar  (\ref{R1}) reduces to
\be
R = \omega \frac{\phi^{,\alpha} \phi_{,\alpha}}{\phi^{2}}
+\frac{3\Box{\phi}}{\phi} - \frac{2V}{\phi}.
\label{R3}
\ee
Now we replace $\Box{\phi}$ from equation (\ref{we1}) and use the
transformations (\ref{trans1}), (\ref{trans2}) and (\ref{bd1}) to get 
$R$ in terms of the variables in the transformed version as 

\be
R=e^{\frac{\psi}{a}}\left ( \frac{\omega}{a^{2}}\psi^{,\alpha}\psi_{,\alpha} - 
2\bar{V}-\frac{3}{a}\frac{d\bar{V}}{d\psi}\right )
\label{R4}
\ee
which in terms of $\bar{R}$ reads like

\be
R = e^{\frac{\psi}{a}}\left[ \frac{2\omega}{a^{2}} \bar{R}
+ 2\bar{V}\left(\frac{2\omega}{a^2}-1\right)
-\frac{3}{a}\frac{d\bar{V}}{d\psi}\right] ,
\label{R5}
\ee
where $a=\sqrt{2\omega+3}$. So it is expected that if $\psi$ and the 
derivative of the potential $\bar{V}$ are regular at the horizon the
regularity features of the curvature scalar are the same for $R$ and
$\bar{R}$. 

\section{a specific example}

Sudarsky \cite{sudarsky} proved a very useful theorem showing that an 
asymptotically flat spherical black hole cannot support a Higgs type
scalar hair. The relevant action is 
\be
S = \int{d^{4}x \sqrt{-g}[\bar{R} -\frac{1}{2}\psi^{,\alpha}\psi_{,\alpha} -\bar{V}(\psi)]},
\label{action3}
\ee
in units where $8\pi G=1$. This action can be readily identified with that 
given by (\ref{action2}) without the gauge field $F_{\mu\nu}$.  With a
very general spherically symmetric metric which is asymptotically flat,
Sudarsky showed that there will be a regular horizon only if the
scalar field is trivial, i.e., $\psi =constant$, and $\bar{V} =0$. Now if
we take a Brans--Dicke action with a potential $V(\phi)$, the
corresponding metric tensor components can be derived with the help of the
inverse transformations using  equation (\ref{trans1}). As these two
spacetimes are conformally related, the site of the horizon would be the
same. Now  Sudarsky's result shows that $\bar{R}$ will not be well
behaved unless $\psi=constant$ and $\bar{V}=0$,which clearly shows
from equations (~\ref{trans2}), (~\ref{pot1}) that the scalar $R$ as given by  
(~\ref{R5})  will be well behaved only if $\phi=constant$  and $V=0$. Thus
a spherical black hole cannot support a Brans--Dicke scalar hair even if
it is endowed with a self-interaction term. \\

Sudarsky's results are valid for an uncharged asymptotically flat
spherical black hole. Hence in view of his theorem and the general
results obtained in the previous section ( equations (~\ref{R1}
and (~\ref{R2})), it can be concluded that no uncharged
asymptotically flat spherical black hole can have a scalar hair even
if the scalar field is self interacting and nonminimally coupled to 
gravity. 

\section{discussion}

Dicke\cite{dicke} used a conformal transformation of the metric tensor to
rewrite the Brans--Dicke field equations in a form which is similar to
that of a minimally coupled zero mass scalar meson field. Saa\cite{saa}
gave a much more generalized version of the technique with the help of
which the gravitational field equations for practically any nonminimally
coupled scalar tensor theory can written as those for a massless minimally
coupled scalar field. In the present work Saa's method is further
generalized to include a self interaction term of the nonminimally coupled
scalar field and any $U(1)$ gauge field. So from the solutions of a
minimally coupled scalar field, the existence or the absence of a scalar
hair can be predicted for a wide class of scalar tensor theories even
including a potential. The non existence of a scalar hair for a self
interacting scalar field was proved by Mayo and Bekenstein\cite{mayo} for
a particular type of coupling between the scalar field and the curvature.
The method developed in the present work can be used for any kind of
coupling. The method will fail in two cases. One is where $f(\phi)$ is
negative, as the method depends crucially on the positivity of $f(\phi)$.
But this perhaps should not be considered as a serious limitaion as in the
weak field limit, $f$ gives the inverse of the Newtonian gravitational
constant, which is expected to be positive. It should be noted that
although the method given in ref.\cite{mayo} works  where the coupling
between $R$ and scalar field is quadratic in $\phi$, the corresponding
$f(\phi)$ need not be restricted to positive values only. The second kind
of coupling which escapes this method are the ones where the matter
Lagrangian is also nonminimally coupled to the scalar field. The example
being the dilaton gravity. \\
The inclusion of the electromagnetic field is also important, as in some
cases where a scalar hair cannot grow independently but it may appear as
a secondary one, i.e., growing on another hair like the electric charge.
Once again the example is the dialton black hole\cite{garfinkle}.

\section{Acknowledgement}
N.B. wishes to thank the Abdus Salam International Centre for Theoretical
Physics for the warm hospitality, where the major part of the work was
done. The work was formulated when N.B. and N.D were visiting the Deapartment
of Theoretical Physics, the University of the Basque Country, Bilbao.
It is a pleasure to thank their hosts Alberto Chamorro and Jose
Senovilla.


\end{document}